 \documentclass[final,5p,times,twocolumn,authoryear]{elsarticle}

\usepackage{amssymb}
\usepackage{lipsum}
\usepackage{amsmath}
\usepackage{natbib}

\newcommand{\aap}{    {\it Astron. Astrophys.}}
\newcommand{\aaps}{   {\it Astron. Astrophys. Suppl.}}

\newcommand{\aj}{     {\it Astron. J.}} 
\newcommand{\apj}{    {\it Astrophys. J.}}

\newcommand{\mnras}{  {\it Mon. Not. Roy. Astron. Soc.}}

\allowdisplaybreaks

\usepackage{color}

\journal{High Energy Astrophysics}

\begin{document}

\begin{frontmatter}



\title{Multifrequency evolution of the Integrated pulse profile of radio pulsars by implementing the inverse Compton mechanism}


\author[first]{Tridib Roy}
\affiliation[first]{organization={Nicolaus Copernicus Astronomical Center},%Department and Organization
            addressline={Rabianska 8, Torun 87-100, Poland}, 
            city={Torun},
            postcode={87-100}, 
            state={},
            country={Poland}}
    
           \author[second]{Mayuresh Surnis}
\affiliation[second]{organization={Indian Institute of science and Education Research, Bhopal},%Department and Organization
            addressline={Bhopal Bypass Road, Bhahuri}, 
            city={Bhopal},
            postcode={462066}, 
            state={Madhya Pradesh},
            country={India}}
 
\author[third]{Mageshwaran Tamilan}
\affiliation[third]{organization={Manipal Centre for Natural Sciences, Manipal Academy of Higher Education},%Department and Organization
            addressline={}, 
            city={Manipal},
            postcode={576104}, 
            state={Karnataka},
            country={India}}

             \author[fourth]{Monalisa Halder}
\affiliation[fourth]{organization={National Institute of Technology Durgapur},%Department and Organization
            addressline={Mahatma Gandhi Avenue}, 
            city={Durgapur},
            postcode={713209}, 
            state={West Bengal},
            country={India}}

\author[fifth]{Siddhartha Biswas}
\affiliation[fifth]{organization={SN Bose National Center For Basic Science},%Department and Organization
            addressline={Salt lake, Sector 3, JD Block}, 
            city={Kolkata},
            postcode={700106}, 
            state={West Bengal},
            country={India}}

\begin{abstract}
The Main Aim of this paper is to explain the emergence of new components of pulsars at higher radio bands by implementing the Inverse Compton Scattering Mechanism.
From pulsar radio observation, it is seen that a couple of pulsars reveal new emission components at higher radio frequencies, although they show single-component emission at lower frequencies. We develop a brief outline, fostering inverse Compton scattering (ICS) of the low-frequency radio photons as a vulnerable source of scattering, susceptible to explaining the evolution of new components of some radio pulsars at higher bands. We couple the conventional curvature radiation (CR) mechanism and ICS, and suggest that the spectral convolution of the flux component individually from CR and the modulated template due to the ICS scattered component can be combined to reproduce such signatures associated with the diverse morphology of the integrated pulse profile. We reproduce the beam frequency diagram, the geometrical variation of different parameters of the emission geometry, as well as the multi-frequency evolution from theory. We have suitably tuned the input parameter space and given the combination of parameters that can tune to a particular scattered frequency in tabulated form.
We conclude that ICS may be a responsible process for describing the emergence of new components in higher radio emission bands.
\end{abstract}



\begin{keyword}
%% keywords here, in the form: keyword \sep keyword, up to a maximum of 6 keywords
Neutron star \sep emission mechanism \sep Inverse Compton Scattering 



\end{keyword}

\end{frontmatter}

%\tableofcontents



\section{Introduction}
\label{introduction}

The magnetosphere of a radio pulsar involves a complex plasma astrophysical process, dealing with the details of the kinematical behavior of primary plasma and its reduction to secondary electron-positron plasma pairs. A magnetic field of about $\sim 10^{12}$ G, together with the pulsar’s rapid rotation, creates a huge potential drop across the polar cap. This potential is strong enough to pull charged particles from the polar gap and accelerate them. In such an extreme magnetic field, any momentum the particles have perpendicular to the field lines reduces to the lowest Landau level as the synchrotron cooling time is very short. Consequently, the primary particles speed up along the magnetic field and emit high-energy photons. These photons, in turn, produce electron–positron pairs, building up a dense cloud of secondary plasma. When this pair cloud becomes dense enough to screen the gap, it triggers coherent processes that give rise to the strong radio emission we observe from pulsars \citep{1975ApJ...196...51R}.

Radio pulsar pulse profiles exhibit a set of standard morphologies, first systematically classified by Rankin. In Rankin’s framework, the radio beam is produced by a concentric arrangement of plasma emission regions that “levitate” above the polar cap. The observed pulse shape is then determined by how the observer’s line of sight slices through this structured beam. A different class of models, following \citet{1981AJ.....86.1953M}, does not assume such an ordered structure. Instead, the emission sources are taken to be distributed more arbitrarily across the beam. Depending on how the line of sight intersects this distribution, pulse profiles with different numbers and shapes of components can appear. However, to understand complete pulse profile morphology, it is necessary to understand emission geometry \citep{2001ApJ...555...31G} and its polarization signature with high-resolution datasets \citep{1999A&A...342..474G}. Statistical interpretation of pulsar data is also very important to understand the nature of its emission. Recent data-based studies by \citet{2024ApJ...973...56C,2025A&A...695A.203J,2015ApJ...800...76S} address different key properties associated with flux fluctuation, nulling, frequency evolution, polar cap geometry, and origin of magnetar's radio emission. 

Pulsar radio emission is broadband, extending from a few MHz to several GHz, and its pulse profiles fall into a set of characteristic morphological types. Types I(a,b) are core-dominated, while Types II(a,b,c) are conal-dominated \citep{2001A&A...377..964Q}. Type I(a) pulsars have very small impact angles and show a single core component at low frequencies, but at higher frequencies the line of sight intersects the inner conal branch, producing a triple profile; for example, PSR B1933+16 is core-single below 1.4 GHz and triple at higher frequencies \citep{1988MNRAS.234..477L,1997A&A...321..519S}. In our current model, we have only attempted to simulate the pulse profile of Type 1a, with a detailed mathematical formulation. Pulsars with Type Ib generally are characterized by a higher line of sight impact angle. At low frequency, they exhibit a core-triple profile, but they show a conal double profile as frequency progresses to a higher value. An exemplary example of this type is PSR B1845-01 \citep{1994A&AS..107..527K}. The conal-dominated group includes Type II(a), in which pulsars can show up to five components at typical frequencies, although at low frequencies, small impact angles result in only three components; PSR B1237+25 is a classic example, with five components merging into three around 50 MHz \citep{1992ApJ...385..273P}. Type II(b) pulsars have somewhat larger impact angles, showing three components at low frequencies and four at higher frequencies when the core branch is no longer intersected, as in PSR B2045-16 \citep{1994A&AS..107..527K}. Finally, Type II(c) pulsars have the largest impact angles, such that only the outer conal branch is intersected, producing conal-double profiles at all frequencies, with the component separation decreasing at higher frequencies; PSR B0525+21 is a representative case \citep{1992ApJ...385..273P}. However, the main aim of this paper is to understand the beaming diagram of pulsar morphology and generate the integrated pulse profile to explain the Type 1a category profile with a detailed mathematical framework.

Because the magnetosphere contains a high number density of charged particles, the plasma frequency is large, which normally prevents low-frequency radio waves from escaping. Some studies suggest that such radiation can escape through a magnetospheric vacuum-like pocket, which is induced by huge plasma radiation-induced rarefication, allowing low-frequency photons to propagate long enough to undergo inverse Compton scattering (ICS). Radio emission can escape primarily due to spatial inhomogeneity and anisotropy in the magnetospheric plasma. Pair cascades in the inner magnetosphere are known to produce strong density fluctuations, including charge-depleted regions associated with intermittent sparking in the polar cap. Such intermittency leads to the formation of low-density “pockets” or tubes along open field lines. In these regions, the effective plasma frequency is significantly reduced compared to the mean Goldreich–Julian density, permitting propagation of electromagnetic modes above the local cutoff. Basically, vacuum-like pockets mean very low density plasma channels or tubes, existing over certain directions at some specific regions in the magnetosphere, such plasma plasma-depleted regions may be created by radiation, plasma pressure-induced rarefication \citep{1996ApJ...460..163S}. However, there are no conclusive remarks or evidence from pulsar observation; these are hypotheses or speculation from a theoretical point of view. Such charge-starved zones or channels where plasma density drops significantly,  due to temporal gap breakdown, inhibition of plasma flows, or pair creation and non-stationary pair creation. These structures may arise naturally in space-charge-limited flow models and partially screened gap models, where the accelerating potential fluctuates and produces bursts of pairs rather than a steady plasma flow.

This motivates applying the ICS mechanism to explain the rich frequency-dependent evolution of pulsar pulse profiles. We suggest that low-frequency photons preferentially scatter off secondary particles with modest Lorentz factors ($\gamma \sim 10$) before they would otherwise be absorbed. The principal mathematical foundation of ICS and its application on radio pulsars was laid down in the classical paper by \citet{1998A&A...333..172Q}. Although our comprehensive understanding of pulse profile morphology was quite rudimentary at that time, subsequent simulation by \citet{2000ApJ...535..354X,1997ApJ...491..891Z,2001A&A...377..964Q,2011ApJ...741....2L} made the basic foundation very clear, giving a comprehensive understanding of the constraints of the ICS parameters, with emission geometry, gap parameter estimation, energy loss of Lorentz factor and diverse pulse profile simulation. Some recent pioneering work shows an unprecedented level of success, fostering the application of coherent ICS mechanisms on Fast radio burst \citep{2024ApJ...972..124Q,2022ApJ...926...73Z} and frequency evolution study with a broader emphasis lying on ICS \citep{2022ApJ...926...73Z}.

While ICS may help account for the emergence of new components at higher observing bands, the precise role of propagation effects in shaping profile morphology remains uncertain. Pulsar radio waves originate deep within a strongly magnetized, ultra-relativistic plasma outflow \citep{Petrova_2016}, where the medium is birefringent and supports two natural polarization modes (O and X) with distinct refractive indices. As a result, the ray paths diverge: the O-mode refracts more strongly, producing broader profiles than the X-mode \citep{kniazev2025datafastmeerkatsurveys}. Numerical integration of the ray path in the pulsar medium with a detailed mathematical treatment shows the frequency-dependent behaviour and polarization property \citep{2010MNRAS.403..569W}. However, ray evolution and propagation effect do not completely alter the property of the output radio emission; rather, it depends upon the plasma-based emission mechanism \citep{2021MNRAS.500.4549M}. Birefringence can also shift the modes in phase, leading to orthogonal polarization jumps, and can yield distinct sub-beams for the two modes so that an observer may see two intensity components at different pulse phases if the modes decouple. Additionally, the modes may couple coherently (e.g. via generalized Faraday rotation) as the plasma conditions evolve, altering the observed polarization states \citep{10.1093/mnras/stad2271}. These effects can dramatically reshape pulse profiles—broadening or splitting components, changing their phase, and altering their number. For instance, detailed modeling shows that strong O-mode refraction in a pair plasma can transform a single emission ring into a three-peaked (“core + cones”) profile \citep{beskin2023triplepulsarprofilesgenerated}. Thus, magnetospheric propagation must be taken seriously, as it can significantly alter the observed morphology relative to that predicted by geometric models alone.

The structure of the paper is as follows. In Section 2, we present the mathematical formulation of the ICS mechanism and our method for simulating the integrated pulse profile. Section 3 provides the results and their interpretation. Sections 4 and 5 contain the discussion and the conclusions, respectively.
\
 \begin{figure*}[h!]
   \centering
   \includegraphics[height=20 truecm, width=18 truecm]{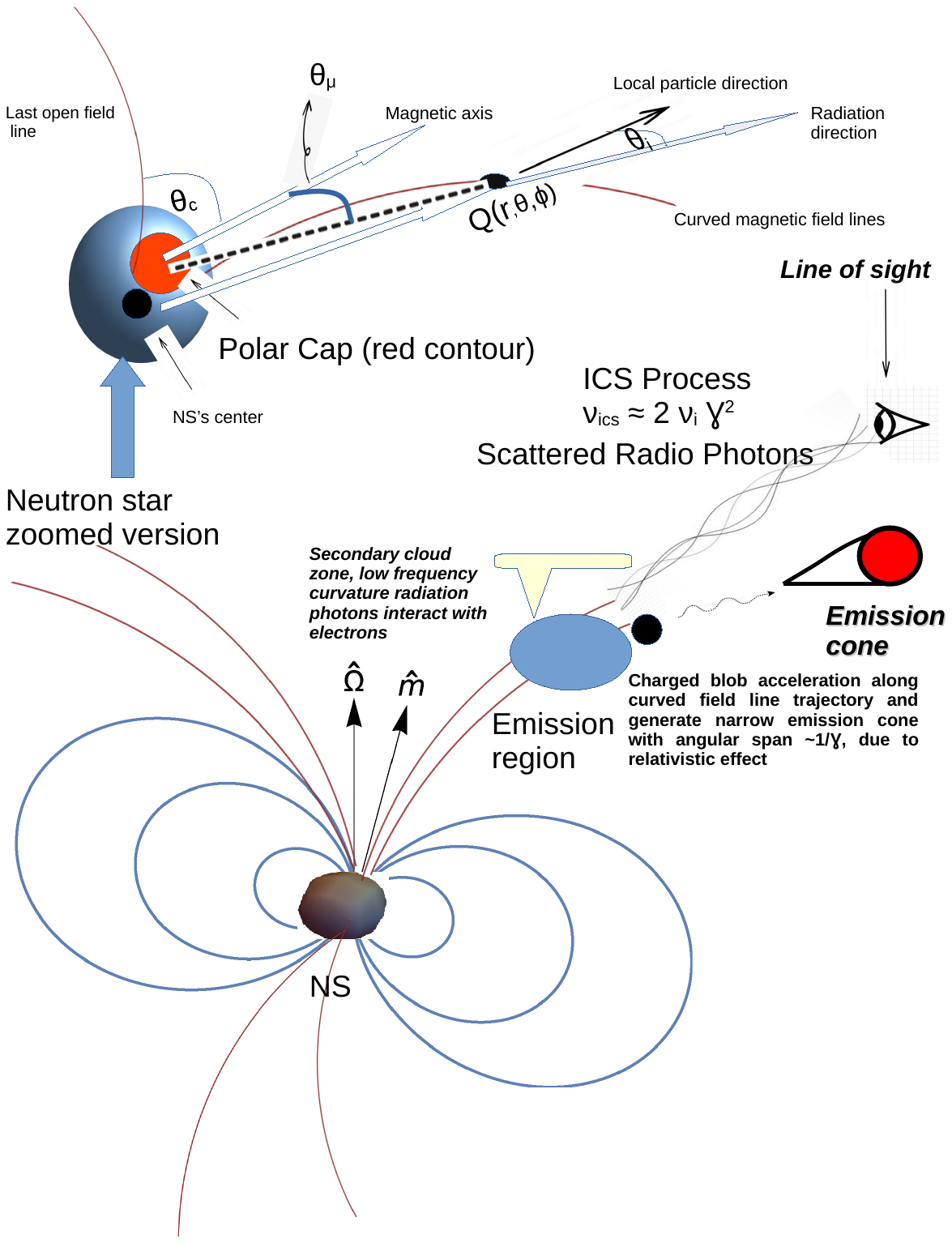}
      \caption{The Above figure shows the overall geometry of the ICS process in the pulsar magnetosphere in the form of a Cartoon diagram. Bottom line picture shows the neutron star embedded in a dipolar structure. $\hat{\Omega}$, $\hat{m}$ represent the the rotation axis and the magnetic moment. ICS process, scattered frequency $\nu_{ics}$ is proportional to the input radio frequency ($\nu_{i}$) and Lorentz factor $\gamma$ of the secondary plasma. The upper side displays a zoomed version of the neutron star, featuring the polar cap and the parameter spaces involved in the ICS geometry. $\theta_{c}$ denotes the angle between the magnetic axis and the last open field line,$\theta_{\mu}$ denotes the angle between the magnetic axis and the emitted radiation direction, and finally $\theta_{i}$ demarcates the angle between the radiation direction and the local motion of plasma particles.}
         \label{fig1a}
   \end{figure*}

\section{Basic formulation on inverse Compton mechanism}

%\subsection{Subsection title}
We consider a low-frequency radio photon that is scattered by the secondary pair plasma in the upper magnetosphere through the inverse Compton mechanism. Figure \ref{fig1a} provides a schematic overview of the current model. Here, $\theta_{\mu}$ denotes the angle between the radiation direction and the magnetic axis, $\theta_{i}$ is the angle between the radiation direction and the local direction of motion of the secondary electrons, and $\theta_{c}$ represents the angular span of the polar cap. The bottom panel of Figure 1 shows a neutron star with its dipolar magnetic field lines. Primary particles flow outward along these lines and, within a few neutron-star radii, undergo pair-cascade discharges that produce a secondary electron–positron plasma. In Figure 1, we mark with a blue ellipse the region where low-frequency photons are scattered to higher radio frequencies before their energy is attenuated. These scattered photons are then convolved with the intrinsic curvature-radiation spectrum and form conal radio beams on either side of the core component in the integrated pulse profile. In this model, we assume that the core component in the higher radio band has an intrinsic origin, arising directly from curvature radiation. The frequency-shifted component produced via the inverse-Compton mechanism adds to this intrinsic emission, yielding the full integrated profile at higher radio frequencies. At lower radio frequencies, however, the profile remains dominated by the intrinsic single core component generated by curvature radiation alone.

If an incoming photon of frequency $\nu_0$ is scattered by the secondary plasma with Lorentz factor $\gamma$, the frequency of the upscattered photon is given by
\begin{equation}\label{eq1}
\nu_{ics}=2\gamma^{2}\nu_{0}(1-\beta\cos\theta_{i}),
\end{equation}
where $\theta_{i}$ is the angle between the local direction of motion of the secondary cloud particles and the incoming photons, such that 
\begin{equation}\label{eq1a}
\begin{split}
\cos\theta_i =& \frac{\mathcal{N}}{\mathcal{D}}, \\[4pt]
\mathcal{N} =& 2 r\cos\theta 
   - R\big[\,3\cos\theta\sin\theta\sin\theta_{c}\cos(\phi-\phi_{c})
    \\ 
    &+ (3\cos^{2}\theta-1)\cos\theta_{c}\,\big],\\[4pt]
\mathcal{D} =& (1+3\cos^{2}\theta)^{1/2} \times \\
 & (r^{2}+R^{2}-2 r R(\cos\theta\cos\theta_{c}+\sin\theta\sin\theta_{c}\cos(\phi-\phi_{c})))^{1/2}.
\end{split}
\end{equation}
Here, $r$ is the radial coordinate of the scattering point or emission spot, where low-frequency photons get converted to a high-energy radio counterpart. $R = R_{\rm Ns}$ is the radius of the neutron star, $ \theta $ is the co-latitude of the emission spot, $\phi$ is the magnetic azimuth, $\phi_{c}$ is the spark location on the polar cap rim. $\theta_{c} = \left( 2\pi R / P c \right)^{1/2}$ is the co-latitude of the last open field line, i.e., the angular span of the polar cap region. Here, $P$ is the spin period of the pulsar, and $c$ is the light velocity.

%Here, $r$ is the radial coordinate of the scattering point or emission spot, where low-frequency photons get converted to a high-energy radio counterpart. $R=R_{Ns}$ is the radius of the neutron star, $\theta$ is the co-latitude of the emission spot, $\phi$ is the magnetic azimuth , $\phi_{c}$ is the spark location on the polar cap rim. $\theta_{c}=(\frac{2\pi R}{P c})^{1/2}$ is the co-latitude of the last open field line, i.e., the angular span of the polar cap region. Here, P is the spin period of the pulsar, and c is the light velocity.

%\begin{equation}\label{eq1a}
%\cos\theta_{i}=\frac{2 r\cos\theta-R(3\cos\theta\sin\theta\sin\theta_{c}\cos(\phi-\phi_{c})+(3\cos^{2}\theta-1))\cos\theta_{c}}{(1+3\cos^{2}\theta)^{1/2}(r^{2}+R^{2}-2 r R(\cos\theta\cos\theta_{c}+\sin\theta\sin\theta_{c}\cos(\phi-\phi_{c})))^{1/2}} 
%\end{equation}

Next, we assume that our model strongly supports curvature radiation and that radio radiation is emitted at the plasma frequency. The plasma frequency expression is given by
\begin{equation}\label{eq2}
\omega_{pe}=\sqrt{\frac{4\pi n_{e} q^{2}}{m_{e}}},
\end{equation}
where $n_{e}$ is the plasma number density, $m_{e}$ is the mass of an electron, and $q$ is the charge of an electron. In strongly magnetized pair plasmas, collective plasma processes regulate the spectrum of the escaping radiation. Even if single-particle curvature radiation is initially generated at its characteristic frequency, propagation effects, mode coupling, and induced scattering in the magnetosphere tend to select frequencies near the local plasma frequency where refractive effects allow more efficient escape. Thus, the emergent radio emission often reflects the plasma frequency of the emitting region rather than the primary curvature emission frequency. We now emphasize that the plasma frequency sets a characteristic scale for escaping radiation, while the curvature mechanism provides the primary energy input and determines the full spectrum of the radio emissions, sets the upper and lower limits of emitted frequency for a given constraint of emission geometry. Mathematically General theory of coherent curvature radiation shows that for efficient transfer of energy to the escaping mode, resonant interaction must take place close to plasma frequency, otherwise output radio emission will diminish (See classic paper \citet{10.1093/mnras/177.1.109} and simulation of  \citet{2021ApJ...911..152G}).

%Now we shift our focus to the context of the ICS paradigm. The current model assumes that the radio emission is excited by the curvature radiation, and the high-energy radio counterpart emerges due to the ICS-mediated scattering of the low-energy radio photons. So essentially low low-energy photons are generated by curvature radiation, and they propagate up to a certain distance and get scattered by the ICS mechanism and generate an extra conal component at higher radio bands in addition to the core. The process is complex as the core component corresponds to a specific radio band, and it's convolved with the ICS scattered low-frequency photon counterpart, and makes the integrated pulse profile. However, the ICS process should operate on the low-frequency photons before they get completely attenuated. 

Now we shift our focus to the ICS framework. In this model, the radio emission is first generated by curvature radiation, producing low-frequency photons that propagate outward through the magnetosphere. As these photons travel to higher altitudes, they are scattered by secondary charges via the inverse Compton mechanism, giving rise to an additional conal component at higher radio frequencies alongside the core emission. The formation of the integrated pulse profile is therefore a convolution of the core component associated with a particular radio band and the ICS-scattered counterpart of the low-frequency photons. For the mechanism to operate effectively, the scattering must occur before the low-frequency photons are fully attenuated. The scattering angle of photons for any multipole-centered star with order $l$ is given as follows (see \citet{2023MNRAS.522.1480D}):
\begin{equation}\label{eq3}
\theta_{sc}=\frac{1}{(l+1)P_{l}(\cos\theta)} \frac{d P_{l}(\cos\theta)}{d\theta} \approx\frac{l}{2}\theta,
\end{equation}
where $P_{l}$ is the Legendre polynomial and $\theta \ll 1$ is considered. For the dipole case ($l=1$), the magnitude of the scattering-induced correction to the polar angle is $\theta/2$. Therefore, after scattering, the net polar angle $\theta_{net}\approx \theta+\theta/2=(3/2)\theta$. 

Fourier component of the radiation electric field at a distance observer can be written as (see \citet{2010ApJ...710...29G}):
\begin{equation}\label{eq4}
E(r,\omega)= A\int_{-\infty}^{\infty}|b|\frac{\hat{n}\times((\hat{n}-\vec{\beta}))\times\dot{\vec{\beta}}) }{\xi^{2}}\exp[i\omega(t-\hat{n}.\vec{r}/c)]d\theta
\end{equation}
where \[ A=\frac{q\exp[i\omega R_{0}/c]}{\sqrt{2\pi} R_{0}\kappa c^{2}}\] and \[\xi=1-\hat{n}.\vec{\beta}\] is the beaming factor. $R_{0}$ is the distance of the pulsar, $\beta$ is the velocity of the particle as a fraction of light velocity, $\dot{\beta}$ is the acceleration factor as a fraction of the light velocity, $\hat{n}$ is the line of sight vector, $|b|$ is the magnitude of the field line trajectory, which is a function of dipolar field line constant $r_e$ and polar angle, $\omega$ is the angular frequency of the emitted mode, which is expected to be close to the plasma frequency of the system. $\kappa$ is the velocity of the plasma blob as a fraction of light velocity. The particle crossing time across the arc length bounded by a given dipolar field line, $t$, is given by:
\begin{equation}\label{eq4a}
t=\frac{1}{\kappa c}\int_{0}^{\theta}|b|d\theta.
\end{equation}
It finally translates to the following expression:
\[
  t=\frac{r_{e}}{12\kappa c}(12+\sqrt{3}\log(14+8\sqrt{3})-3\cos\theta\sqrt{10+6\cos2\theta}-\] \[2\sqrt{3}\log(\sqrt{6}\cos\theta+\sqrt{5+3\cos2\theta}))  
\]
Now, we define two separate terms that arise in the Fourier integral associated with the radiation of the electric field (equation \ref{eq4}). The principal argument is given by:
\begin{equation}\label{eq5}
\mathcal{A}=\frac{|b|}{\kappa c}\frac{\hat{n}\times((\hat{n}-\vec{\beta}))\times\dot{\vec{\beta}}) }{\xi^{2}},
    \end{equation}
 where the magnitude of the tangent vector for a dipolar trajectory is given by \citet{2010ApJ...710...29G}:
 \begin{equation}\label{eq5a}
     |b|=\frac{r_{e}\sin\theta}{\sqrt{2}}(5+3\cos(2\theta))^{1/2}.
 \end{equation}
 %where $r_{e}$ is the dipolar field line constant, and $\theta$ is the polar angle associated with the emission spot.
The exponential argument is given by
\begin{equation}
C_{z} = \omega\left(t-\frac{\hat{n}.\vec{r}}{c}\right).
\end{equation}
We consider these two arguments in detail below.
 
By using spherical geometry and coupling the dipolar field configuration, one can estimate the Dot product of the following quantity as \citep{2010ApJ...710...29G}:
 \begin{multline}\label{eq5b}
\hat{n}\cdot\vec{r}=r_{e}\sin^{2}(\theta)\left[\cos\alpha(\cos\theta\cos\zeta+\cos\phi\cos\phi^{\prime}\sin\theta\sin\zeta)\right.\\ \left. -\cos\zeta\cos\phi\sin\alpha\sin\theta+\sin\zeta(\cos\theta\cos\phi^{\prime}\sin\alpha \right. \\ \left. -\sin\theta\sin\phi\sin\phi^{\prime})\right]
 \end{multline}
The power series expansion of $C_{z}$ around the center of emission point ($\theta_0$) is obtained by substituting  the radial vector expression of accelerated charged particles constrained to move along a dipolar trajectory (see \citet{2010ApJ...710...29G}): 
%Exponential argument $C_{z}$ and its power series expansion around the center of emission point by substituting  the radial vector expression of accelerated charged particles constrained to move along a dipolar trajectory (see Gangadhara 2010) :
   \begin{equation}\label{eq6}
       C_{z}=C_{0}+C_{1}(\theta-\theta_{0})+C_{2}(\theta-\theta_{0})^{2}+C_{3}(\theta-\theta_{0})^{3},
   \end{equation}
   where $C_{0},C_{1},C_{2},C_{3}$ are all functions of the emission geometry of pulsars, which are numerically computed in our model to simulate the integrated pulse profile.
   Now we separate the power-series terms that arise from the three vector components of the principal argument as follows:
   \begin{equation}
       \mathcal{A}_{x}=\mathcal{A}_{x0}+\mathcal{A}_{x1}(\theta-\theta_{0})+\mathcal{A}_{x2}(\theta-\theta_{0})^{2}+\mathcal{A}_{x3}(\theta-\theta_{0})^{3}
   \end{equation}

   \begin{equation}
       \mathcal{A}_{y}=\mathcal{A}_{y0}+\mathcal{A}_{y1}(\theta-\theta_{0})+\mathcal{A}_{y2}(\theta-\theta_{0})^{2}+\mathcal{A}_{y3}(\theta-\theta_{0})^{3}
   \end{equation}

   \begin{equation}
       \mathcal{A}_{z}=\mathcal{A}_{z0}+\mathcal{A}_{z1}(\theta-\theta_{0})+\mathcal{A}_{z2}(\theta-\theta_{0})^{2}+\mathcal{A}_{z3}(\theta-\theta_{0})^{3}
   \end{equation}
Therefore, the radiation electric field expression given by equation (\ref{eq4}) translates as:
\begin{multline}
    E_{x}(\omega)=E_{0}\int_{-\infty}^{\infty}(\mathcal{A}_{x0}+\mathcal{A}_{x1}\mu+\mathcal{A}_{x2}\mu^{2}+\mathcal{A}_{x3}\mu^{3})\\\exp[i(C_{1}\mu+C_{2}\mu^{2}+C_{3}\mu^{3})]d\mu,
\end{multline}
\begin{multline}
    E_{y}(\omega)=E_{0}\int_{-\infty}^{\infty}(\mathcal{A}_{y0}+\mathcal{A}_{y1}\mu+\mathcal{A}_{y2}\mu^{2}+\mathcal{A}_{y3}\mu^{3})\\ \exp[i(C_{1}\mu+C_{2}\mu^{2}+C_{3}\mu^{3})]d\mu,
\end{multline}
\begin{multline}
    E_{z}(\omega)=E_{0}\int_{-\infty}^{\infty}(\mathcal{A}_{z0}+\mathcal{A}_{z1}\mu+\mathcal{A}_{z2}\mu^{2}+\mathcal{A}_{z3}\mu^{3})\\ \exp[i(C_{1}\mu+C_{2}\mu^{2}+C_{3}\mu^{3})]d\mu,
\end{multline}
where $\mu = \theta-\theta_0$ and $E_0 = (q/\sqrt{2\pi}R_{0} c)\exp[i(C_{0}+(\omega R_{0}/c))]$. The integral involved in the above set of equations has a specific solution; therefore, the above set of equations reduces to the following:
\begin{equation}
    E_{x}(\omega)=E_{0} (\mathcal{A}_{x0}S_{0}+\mathcal{A}_{x1}S_{1}+\mathcal{A}_{x2}S_{2}+\mathcal{A}_{x3}S_{3})
\end{equation}
\begin{equation}
    E_{y}(\omega)=E_{0} (\mathcal{A}_{y0}S_{0}+\mathcal{A}_{y1}S_{1}+\mathcal{A}_{y2}S_{2}+\mathcal{A}_{y3}S_{3})
\end{equation}
\begin{equation}
    E_{z}(\omega)=E_{0} (\mathcal{A}_{z0}S_{0}+\mathcal{A}_{z1}S_{1}+\mathcal{A}_{z2}S_{2}+\mathcal{A}_{z3}S_{3})
\end{equation}
where the various components of $\mathcal{A}$ in the above equations are geometric functions of dipolar tilt angle $\alpha$, rotation phase $\phi^{\prime}$, line of sight impact angle $\sigma$, dipolar field line constant $r_{e}$. For a given geometry, we computed all power series coefficients numerically. The $S_0$, $S_1$, $S_2$ and $S_3$ are given by
\begin{equation}
S_{0}=\int_{-\infty}^{\infty}\exp[i(C_{1}\mu+C_{2}\mu^{2}+C_{3}\mu^{3})]d\mu = U j_0,
\end{equation}

\begin{eqnarray}\label{Airy2}
S_{1}=&&\int_{-\infty}^{\infty}\mu \exp[i(C_{1}\mu+C_{2}\mu^{2}+C_{3}\mu^{3}]d\mu \nonumber \\ 
=&&\frac{U}{C_{3}^{1/3}}\left(j_{1}-j_{0}\frac{C_{2}}{3 C_{3}^{2/3}}\right),
\end{eqnarray}

\begin{eqnarray}\label{Airy3}
S_{2}=&&\int_{-\infty}^{\infty}\mu^{2} \exp[i(C_{1}\mu+C_{2}\mu^{2}+C_{3}\mu^{3}]d\mu \nonumber \\
 =&& \frac{U}{3 C_{3}}\left[\left(\frac{2 C_{2}^{2}}{3 C_{3}}-C_{1}\right)j_{0}-\frac{2 C_{2}}{3 C_{3}^{1/3}}j_{1}\right],
\end{eqnarray}

\begin{eqnarray}\label{Airy4}
S_{3}=&&\int_{-\infty}^{\infty}\mu^{3} \exp[i(C_{1}\mu+C_{2}\mu^{2}+C_{3}\mu^{3}]d\mu \nonumber \\
 =&&\frac{U}{9 C_{3}^{7/3}}\Bigg[\left(\frac{9 C_{1}C_{2}C_{3}-4 C_{2}^{3}+9i C_{3}^{2}}{3 C_{3}^{2/3}}\right)j_{0} \nonumber \\ 
 &&  + (4 C_{2}^{2}-3 C_{1}C_{3})j_{1}\Bigg],
\end{eqnarray}

where 

\begin{eqnarray}
    U=&&\frac{1}{C_{3}^{1/3}}\exp\left[i \frac{C_{2}}{3 C_{3}}\left(\frac{2 C_{2}^{2}}{9 C_{3}}-C_{1}\right)\right]  \\
    j_{0}=&& \frac{2\pi}{3^{1/3}}Ai\left(\frac{z}{3^{1/3}}\right)  \\
    j_{1}=&& -\frac{2\pi i}{3^{2/3}}Ai^{\prime}\left(\frac{z}{3^{1/3}} \right) \\
    z =&& \frac{1}{C_{3}^{1/3}}\left(C_{1}-\frac{C_{2}^{2}}{3 C_{3}}\right),
\end{eqnarray}
with $Ai(z)$ being the Airy function of $z$ with no branch cut singularity and $Ai^{\prime}(z)$ being the derivative of Airy function $Ai(z)$.

Due to inverse Compton scattering, low-frequency photons are shifted into higher radio emission bands. The scattering outcome depends critically on the intersection point that is, the precise location where the secondary plasma interacts with the incoming low-frequency photons. In principle, this location can vary, since the region occupied by the secondary plasma may extend over several kilometers within the pulsar magnetosphere. The localization of the scattering zone depends on several factors, including the decay profile of the Lorentz factor, the attenuation of low-frequency radio waves in the ambient plasma, and the plasma resistivity, among others.
The complex radiation Electric field vector in the Fourier domain can be written as:
\begin{equation}
E_{net}=\sum_{i=x,y,z}E_{i}(\omega)=E_{x}(\omega)+E_{y}(\omega)+E_{z}(\omega)
\end{equation}
So, two mutually perpendicular components are constructed by segregating components of the electric field in the plane of the sky as:
\begin{equation}
E_{\parallel}=-\cos\zeta E_{x}(\omega)+\sin\zeta E_{z}(\omega)
\end{equation}
\begin{equation}
E_{\perp}=E_{y}(\omega)
\end{equation}
Finally, Stokes intensity or flux is computed as:
\begin{equation}
I=E_{\parallel}E^{*}_{\parallel}+E_{\perp}E^{*}_{\perp}
\end{equation}

%Due to ICS scattering in the higher radio band, it will create a template, where additional spectral components will be generated due to radiation-plasma modulation along polar directions. It implies that the integrated spectrum will be the convolution of the intrinsic curvature radiation component and additional conal components due to the ICS process. So, if the scattering angle due to ICS is $\pm \theta_{sc}$ with respect to the original trajectory, then radiation will be modulated by the sum of two Gaussian functions. If the original central component suffers modulation due to nonuniform plasma, then that factor also needs to be considered.

Due to ICS scattering at higher radio frequencies, the resulting emission forms a template in which additional spectral components arise from radiation–plasma modulation along the polar direction. This implies that the integrated spectrum is effectively the convolution of the intrinsic curvature-radiation component with the additional conal components generated by the ICS process. If the scattering angle due to ICS is $\pm \theta_{sc}$ relative to the original photon trajectory, then the resulting radiation is modulated by the sum of two Gaussian functions. If the original central component suffers modulation due to nonuniform plasma, then that factor also needs to be considered. So, the net Gaussian functions will be:
\begin{equation}
f=f_{t1}\exp[(\frac{(\theta-\theta_{sc})}{\sigma_{\theta}})^{2}]+f_{c}\exp[-(\frac{\theta}{\sigma_{\theta}})^{2}]+f_{l1}\exp[-(\frac{\theta+\theta_{sc}}{\sigma_{\theta}})^{2}]
\end{equation}
where $f_{t1},~f_{c}$ and $f_{l1}$ are the magnitude of the modulation at the trailing location, core location, and at the leading location, respectively.  Due to the generation of an extra component from the lower frequency part by the ICS mechanism, plasma gets modulated along the polar direction, so one needs to constrain it by spectral convolution from two counterparts: (i) intrinsic curvature radiation, (ii) ICS scattered component. Finally, we superpose the modulation template with the original radiation electric field in the Fourier domain and integrate over the emission region across each open magnetic field line for a given pulsar geometry.

The relativistic beaming effect makes the radiation spread over a narrow angular region with respect to the original tangential path. If the emission blob corresponds to a Lorentz factor $\gamma$, the angular span associated with the dimension of the emission cone would scale inversely with the Lorentz factor, i.e., $\gamma^{-1}$. So, integration is done by constructing an emission grid across each contributing field line along the line of sight, i.e., segmenting a narrow cone area and constructing an equivalent area in the plane of polar and azimuthal angle $(\theta-\phi)$ plane by the projection method. So, mathematically net flux is computed from Stokes I as:
\begin{equation}\label{inet}
I_{net}=\int_{\phi_{0}-\delta\phi}^{\phi_{0}+\delta\phi}\int_{\theta_{0}-\delta\theta}^{\theta_{0}+\delta\theta}f I \sin\theta d\theta d\phi
\end{equation}
Here $\delta\theta$ and $\delta\phi$ are the span of the emission region along polar and azimuthal direction due to relativistic beaming effect, whose expression can be found in equation (8) and equation (28) in \citet{2010ApJ...710...29G}. Alternatively, the reader can find emission region computations in Figure 3 of \citet{2025ApJ...994..189R}. The above equation is used to compute and understand the frequency evolution of the flux profile of radio pulsar emission, which combines the modulation parameter "f" and, intrinsic curvature radiation source function. ICS-induced interaction is already included in the simulation code through modulation. In our simulation, the magnitude of the tangent vector of the field line trajectory $|b|$, the radius of curvature of the field line $\rho$ is a function of the field line constant $r_{e}$ and polar angle $\theta$. For some cases, we have treated the dipolar field line constant as a fixed value throughout the entire pulse, but in reality, it's a function of polar angle, emitted frequency ($\nu$), and Lorentz factor ($\gamma$). So, it is expected that the field line constant $r_{e}$ should show phase-dependent variation. The generalized field line constant is given by the following expression (see \citet{2021ApJ...911..152G}):
\begin{equation}\label{flc}
    r_{e}=\frac{9 c\gamma^{3}(3+\cos(2\theta))\csc(\theta)}{2\sqrt{2}\pi(5+3\cos(2\theta))^{3/2}\nu}
\end{equation}
From the above equation, we can see that the emitted frequency $\nu$ as constrained from a given geometry is a function of field line constant $r_e$, Lorentz factor $\gamma$, and polar angle $\theta$. Partial differentiation of the above equation reduces to:
\begin{equation}
d\nu=\frac{\partial \nu}{\partial r_{e}}\delta r_{e}+\frac{\partial \nu}{\partial \gamma}\delta\gamma+\frac{\partial \nu}{\partial \theta}\delta\theta
\end{equation}
where,
\[\frac{\partial \nu}{\partial r_{e}}=-\frac{9 c \gamma^{3}\csc\theta(3+\cos(2\theta))}{2\sqrt{2}\pi r_{e}^{2}(5+3\cos(2\theta))^{3/2}} ,\]\\
\[\frac{\partial \nu}{\partial \gamma}=\frac{27 c \gamma^{2}\csc\theta(3+\cos(2\theta))}{2\sqrt{2}\pi r_{e}(5+3\cos(2\theta))^{3/2}},\]\\
\[\frac{\partial \nu}{\partial \theta}=\frac{-9 c \gamma^{3}\csc^{2}(\theta)(30\cos(\theta)+31\cos(3\theta)+3\cos(5\theta))}{4\sqrt{2}\pi r_{e}(5+3\cos(2\theta))^{5/2}}.\]
If we assume that the pulsar emission arises from a monoenergetic electron distribution, i.e., $\delta\gamma\rightarrow0$, still within a narrow pulse longitude, $\partial \nu / \partial r_{e}$ and $\partial \nu / \partial \theta$ correspond to different values within a particular pulse component. If the core width is large, we can expect some scatter or spread of frequency with respect to the central frequency, as we move towards the tail of a component from the peak locations, because $\partial \nu / \partial \theta$, $\partial \nu / \partial \gamma$, $\theta$ all are functions of rotation phase $\phi^{\prime}$. But for a narrow pulse, frequency scatter would be significantly small. However, it is evident that the conal set of rings corresponds to a different set of field lines, possibly with different $r_{e}$ and $\delta r_{e}$ values. In pragmatic sense, we use $r_{e}$ values to be an integer multiple of the light cylinder radius $r_{Lc}$, which is huge. So it naturally sets the $\partial \nu / \partial r_{e}\rightarrow 0$. So, we expect that frequency spread $\delta\nu$ concerning the central observation frequency arises from $\partial \nu / \partial \gamma$ and $\partial \nu / \partial \theta$ terms. Therefore, pulsar radio emission is perceived to be a broadband phenomenon.

\section{Results and Interpretations}

 \begin{figure*}[h!]
   \centering
   \includegraphics[height=12 truecm, width=18 truecm]{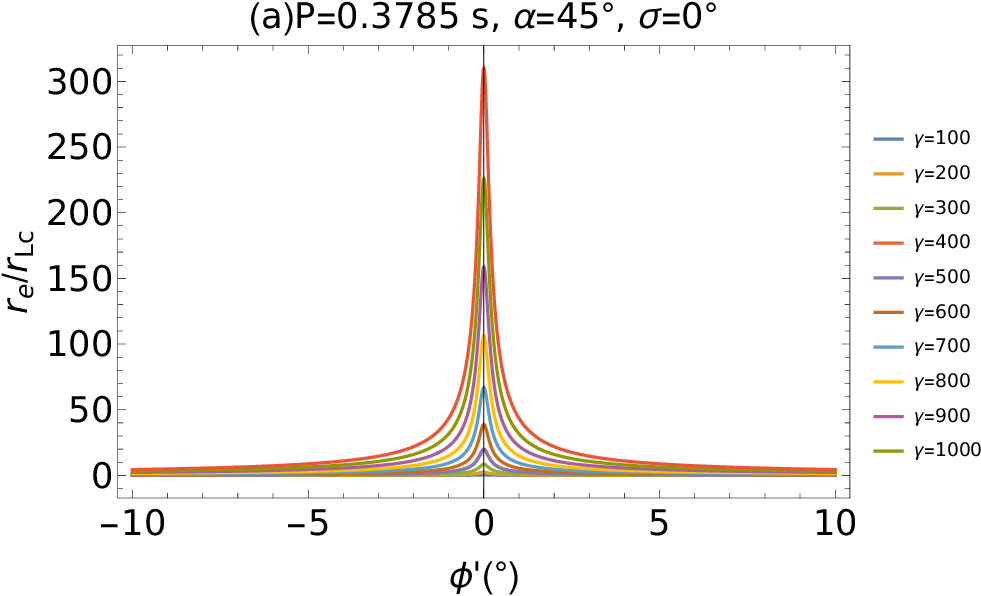}
      \caption{ Figure corresponds to field line constant as a fraction of light cylinder radius vs longitude for different Lorentz factor, we have chosen $\alpha=45^{0}$, $P=0.37 ~$, $\sigma=1^{\circ}$ for the plot. The other parameters are well explained in the text.}
         \label{fig2a}
   \end{figure*}
\begin{figure*}[h!]
   \centering
   \includegraphics[height=12 truecm, width=18 truecm]{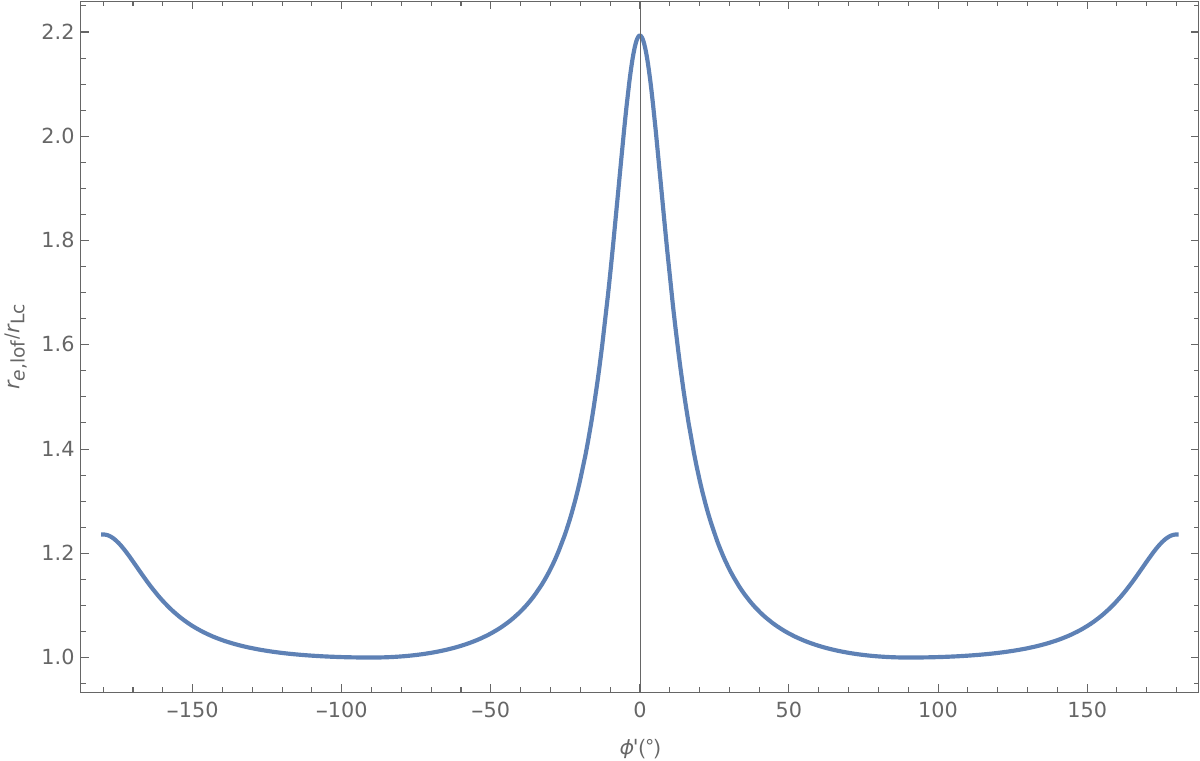}
      \caption{Above figure shows the last open field line constant as a fraction of light cylinder radius for $\alpha=45^{0}$, $P=0.37 ~$, $\sigma=1^{\circ}$.}
         \label{fig2b}
   \end{figure*}

   \begin{figure*}[h!]
   \centering
   \includegraphics[height=12 truecm, width=18 truecm]{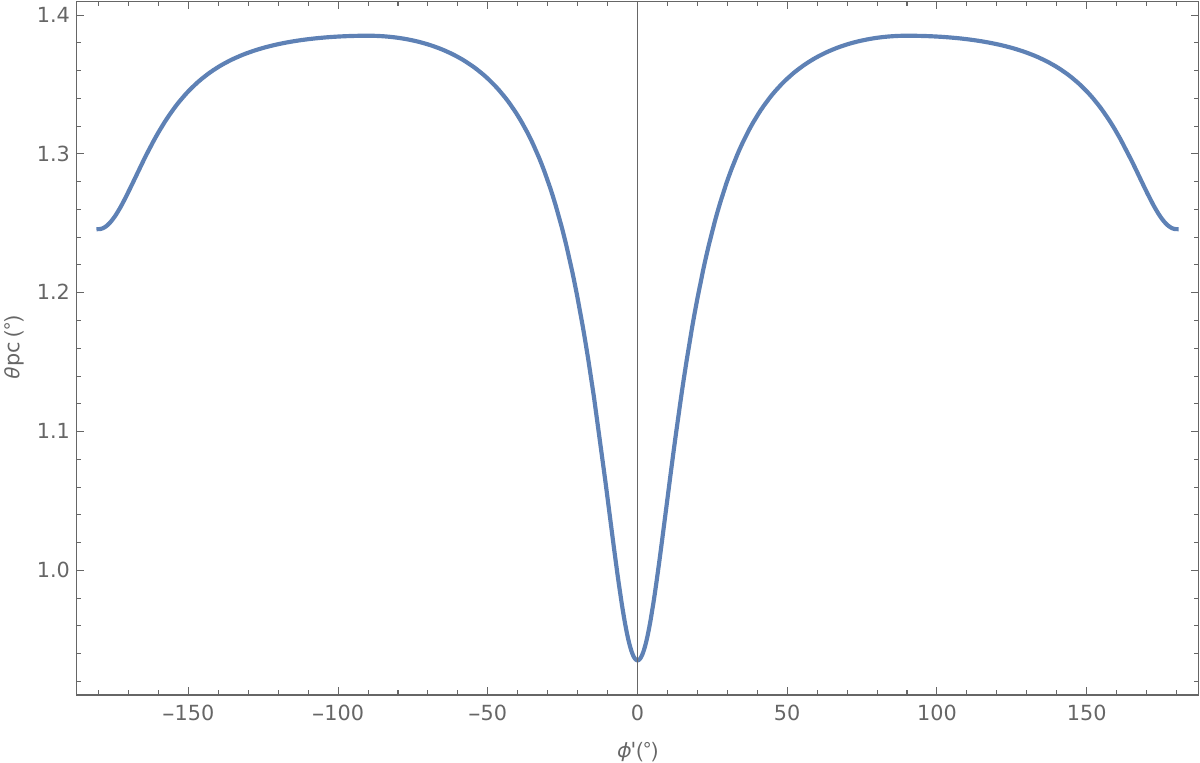}
      \caption{Above figure shows the variation of the polar angle corresponding to the last open field line with respect to rotation phase for $ \alpha=45^{0}$, $P=0.37 ~$, $\sigma=1^{\circ}$. }
         \label{fig2c}
   \end{figure*}
   \begin{figure*}[h!]
   \centering
   \includegraphics[scale = 0.7]{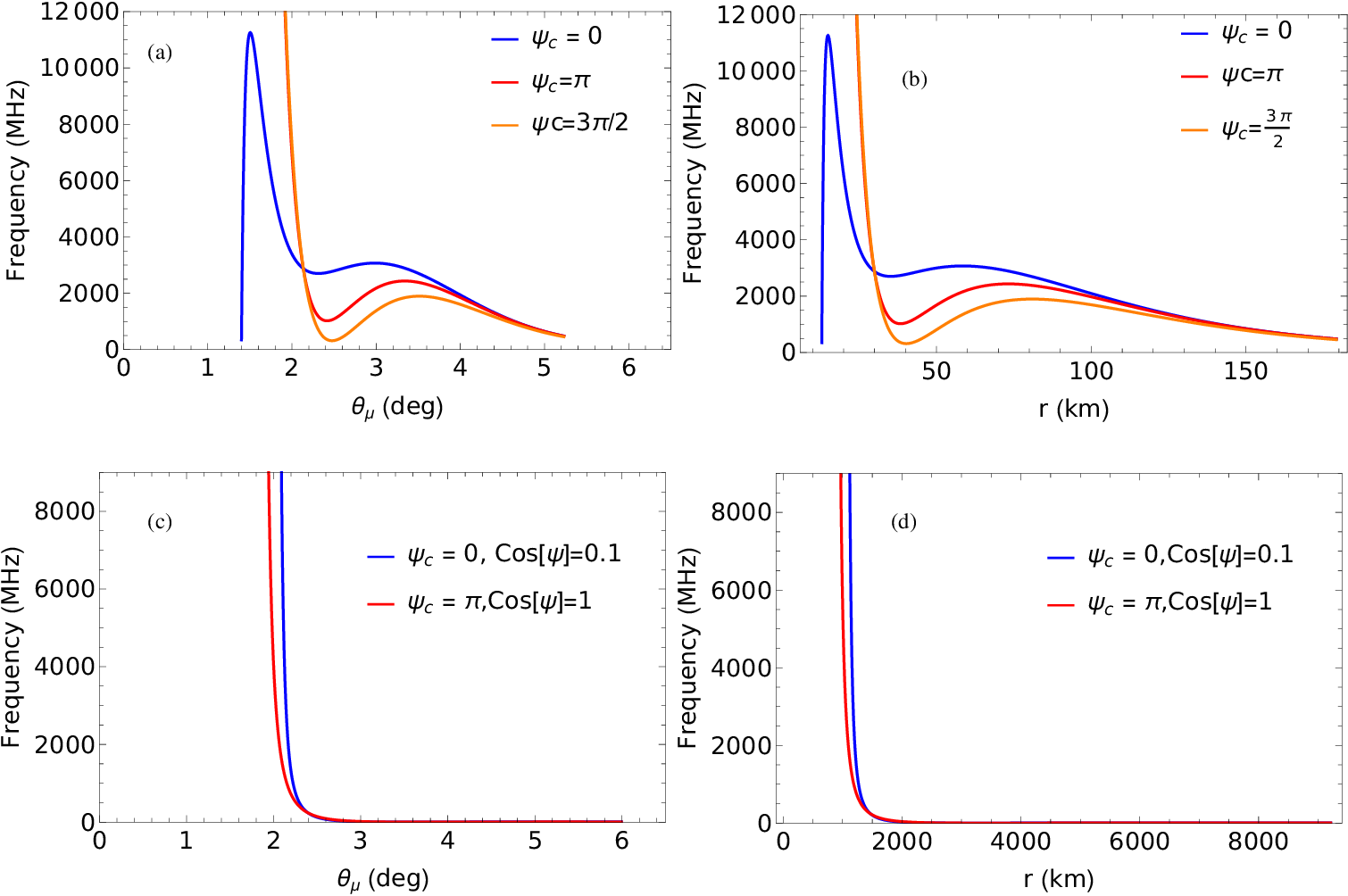}
      \caption{Figures (a) and (b) correspond to the beam frequency diagram for PSR B2111+46, representing the output emitted frequency vs beaming angle and emitted frequency vs emission radii, respectively. Parameter space chosen for Figures (a)-(b) are, spin period of pulsar $P=1$, Lorentz factor $\gamma_{0}=1000$, decay constant $\xi=0.16$, $R_{Ns}=10 ~Km$, input seed photon frequency $\nu_{i}=100~MHz$, dipole tilt angle $\alpha=9^{\circ}$, line of sight impact angle $\sigma=1.4^{\circ}$. Figures (c)-(d) represent the beaming diagram of PSR B1933+16 and to generate it we have chosen $P=0.37~S $, $\gamma_{0}=300$, $\xi=0.16$, $R_{Ns}=10~Km$, $\nu_{i}=1~MHz$. Rest parameters are demarcated in the diagram with a colour legend. To comply with the symbol as used in the paper by \citet{2007A&A...465..525Z}, we have used $\psi$, and $\psi_{c}$ in the legend of the figure, which is equivalent to $\phi$ and $\phi_{c}$ in the current paper, implies $\psi=\phi$ denotes azimuth and $\psi_{c}=\phi_{c}$ marks the azimuthal location of the sparks.}
         \label{fig3a}
   \end{figure*}   

   \begin{figure*}[h!]
   \centering
   \includegraphics[scale = 0.7]{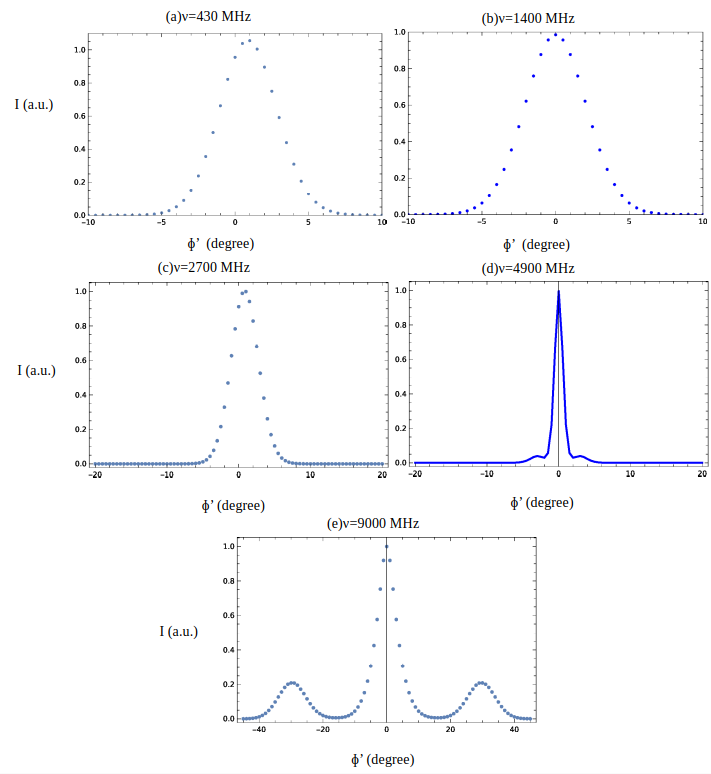}
      \caption{Plots show the simulation of the net flux profile representing I$_{net}$ derived from equation (\ref{inet}), integrated over the emission region for a given Lorentz factor associated with plasma. Simulated pulse profile is mediated by ICS-induced modulation in the polar direction at higher frequency radio band, i.e., 4.9 GHz, 9 GHz emissions. For the generation of multifrequency evolution of flux in the above picture, spanning from 430 MHz to 9 GHz, we use the following input parameters,i.e., spin period P=0.3785 S, dipole axis obliquity parameter $\alpha=44^{\circ}$, line of sight impact angle $\sigma=1^{\circ}$, input frequency of low energy radio photon 1-10 MHz, Lorentz factor of secondary plasma 10 approximately. The y-axis represents Stokes I in arbitrary units, counting the next flux, computed from equation \ref{inet}, and the x-axis shows the rotation phase $\phi^{\prime}$ in degrees.}
         \label{fig4a}
   \end{figure*}

Figure \ref{fig2a} shows the variation of the magnetic field line constant vs rotation phase $\phi^{\prime}$ for different Lorentz factors for a given input frequency $\nu=600$ MHz, which is an approximately symmetric type profile, peaked centrally. We have used equation (\ref{flc}) to plot it's variation in figure \ref{fig2a}. Such a set of curves can be produced for different frequency bands, but they will correspond to a lower or higher peak magnitude at the center point of the diagram, depending on the chosen frequency. Field line constant is inversely proportional to frequency, so generally low frequency emission occurs from field lines with higher field line constant and high frequency emission corresponds to a lower field line constant. At higher rotation phase, the fractional field line constant gradually decreases to a lower value, but the peak magnitude of the field line constant is shifted to a higher value for a higher Lorentz factor for a given radio band, as it's proportional to $\gamma^{3}$. 

Figure \ref{fig2b} represents the last open field line constant as a fraction of the light cylinder radius vs rotation phase, which is computed with the help of equation (15) of \citet{2004ApJ...609..335G} for a given dipole tilt angle $\alpha=45^{\circ}$, line of sight impact angle $\sigma\approx 1$. It shows a reflection of the dipolar field geometry, and depicts a peak value at $\phi^{\prime}=0$ and sharply declines following that. Nevertheless, such variation is purely natural in the context of emission geometry embedded with a global dipolar structure solution \citep{2025ApJ...980...65R}.

Next parameter $\theta_{pc}$ corresponds polar angle corresponds to the last open field line, we have used its expression from equation (13) of \citet{2004ApJ...609..335G}. Figure \ref{fig2c} shows the variation of $\theta_{pc}$ with respect to rotation phase. For $|\phi^{\prime}|\leq50 ^{\circ}$, $\theta_{pc}$ is much less compare to 1 degree but beyond $\phi^{\prime}=\pm50^{\circ}$, it's value shoots up to $1^{\circ}.4$ and roughly remains constant. For the higher rotation phase again it shows a gradual fall. For computing the integrated pulse profile of a radio pulsar, it's important to constrain the emission geometry and understand the field lines morphology, the last open field line constant, and its corresponding polar angle. Such studies help to understand whether radiation corresponds to an open field line or to a closed field line zone.

Figure \ref{fig3a} shows the beam–frequency diagram in both the frequency and spatial (altitude) domains. Panels (a) and (b) display the inverse-Compton–scattered frequency as a function of the beaming angle for PSR B2111+46. The resulting diagram is bounded by three sets of curves, each corresponding to a different spark location on the polar cap. A saddle-point region appears between the crest and trough of these curves. For a fixed emission altitude or frequency band, the beaming diagram clearly divides into three distinct zones. At higher observing frequencies, however, the emission region is bounded by only two curves. Thus, if the observation frequency is sufficiently high, the line of sight intersects only a single component, whereas at lower frequencies a core–cone structure may appear. We also attempted to reproduce the beam–frequency diagram for PSR B1933+16. As shown in panels (c) and (d), the entire radio band is bounded by only two sets of curves for all chosen spark locations; the curves nearly coincide and show no appreciable separation. Therefore, for a fixed emission height, only a single profile component is expected. Since PSR B1933+16 does not exhibit well-separated conal components and is dominated by a single core feature, this beam diagram naturally explains its observed morphology. Although this pulsar develops additional components at higher radio frequencies, these are not intrinsic to the curvature-radiation emission mechanism, which implies that they are not natural components; rather, they are generated by a scattering mechanism. 

In Figure \ref{fig4a}, we have shown a simulation of the frequency evolution characteristics of the integrated pulse profile to explain the component emergence at the higher emission band. We carry out the simulation at 430 MHz, 1.4 GHz, and 2.7 GHz from first principles, i.e., radiation from accelerated charged particles. For the simulation, we have used dipole tilt angle $\alpha = 44^{\circ}$, line-of-sight impact angle $\sigma = 1^{\circ}$, and pulsar spin period $P = 0.3785~\mathrm{s}$. According to our perception, the width of the pulse profile is mostly determined by the input geometrical constraints of the pulsar. At 430 MHz, we can see that the on-pulse width is confined within $\phi' = \pm 5^{\circ}$, whereas at 1.4 GHz the on-pulse width slightly increases, and at 2.7 GHz the width remains almost the same. However, the most crucial behaviour appears at higher radio frequencies, especially for PSR B1933+16, which shows new component generation at 4.9 GHz and close to 9 GHz. As pulsar radio emission is a broadband phenomenon, emission characteristics at some adjacent bands are expected to be the same. However, we conjecture that component emergence is possible in the context of the ICS scenario. Here we consider low-frequency photons at $10$–$100~\mathrm{MHz}$, which propagate up to a certain distance in the magnetosphere and get scattered by secondary particles with a relatively low Lorentz factor $\gamma \sim 10$, before the photons are completely attenuated. Such ICS-scattered beams form conal outer rings on either side of the core. The angular separation between the core and the conal outer rings depends on several factors: (i) the location in the magnetosphere where the low-frequency photons are scattered and the corresponding polar angle of the scattering point.

So, finally, the conal beams convolve with the core beam, which is generated by the intrinsic curvature emission mechanism, and together they shape the integrated pulse profile. To simulate this scenario mathematically, we assume that the ICS scattered photons modulate with the plasma either in the azimuthal or polar direction due to non-uniform plasma distribution and radiation–plasma interaction. From the given peak location and strength of the Gaussian modulation in the azimuthal or polar direction, an equivalent peak location in rotation-phase space can be obtained using the standard radio-emission geometry of pulsars. For convenience in our simulation code, we often convert the peak location of the Gaussian modulation (along $\theta$ direction) to rotation-phase space, and then integrate the emission over the region across each magnetic field line to compute the integrated pulse profile (Stokes $I$).

For simulation at a higher radiation frequency, we need to combine intrinsic curvature radiation photons and the scattered low-frequency photons. To do this, we need to generate the Gaussian modulation template by suitably constraining the input parameter space and combining it with the source photon. Details of the Gaussian templates to generate the integrated pulse at 4.7 GHz and at 9 GHz are shown in Table \ref{Table1}. Next, field line constant values chosen for the simulation at 430 MHz, 1.4 GHz, and 2.7 GHz are shown in Table \ref{Table2}. Table \ref{Table3} shows the details input parameter space, which is susceptible to producing ICS scattered photons at 4.9 GHz and at 8.7 GHz for a given low-frequency photon at 100 MHz. Finally, Table \ref{Table4} shows some combinations of input parameter space, which can generate scattered frequency at 9 GHz.

For the $4.7$ GHz simulation, we superpose a Gaussian modulation in the azimuthal direction and combine it with the source function of intrinsic intensity. In this case, the modulation strength (or magnitude) is chosen to be weaker than that used for the $9$ GHz case. 
For the $9$ GHz simulation, the ICS scattering location is taken to be at a higher altitude, which naturally increases the angular separation between the core and the conal beams. At $9$ GHz, we compute the equivalent rotation phase corresponding to the scattering point. If the scattering occurs at a polar angle $\theta$, then after scattering, the conal beam follows a trajectory with polar angle $1.5 \theta$. By choosing a suitable dissipation factor $\xi$, we identify the location in the magnetosphere where primary plasma with Lorentz factor $\gamma \sim 10^{3} - 10^{6}$ decelerates to a lower value $\gamma \sim 10$. Once such a low-$\gamma$ population is obtained, the corresponding polar angle is derived from the dipolar field-line equation. These lower-Lorentz-factor particles ($\gamma \sim 10$) are capable of producing photons in the higher radio band via the ICS process.

Thus, in the current model, the dissipation factor $\xi$ plays a crucial role in adjusting the angular separation of the conal beams. For the $9$ GHz simulation, we compute the equivalent rotation-phase peak of the Gaussian modulation from its known peak value in the polar direction. We then combine the modulation template with the source function, and finally integrate over the emission region to obtain the behaviour of the integrated pulse profile. One notable aspect is that for the $9$ GHz case we maximize the Gaussian-modulation magnitude, setting it to $1$, which enhances the amplitude of the conal component. We also adopt a smaller dissipation factor $\xi$ for the $9$ GHz case, which forces the Lorentz factor to drop to $\gamma \sim 10$ at a higher altitude, corresponding to larger polar and scattering angles. This naturally increases the separation of the conal components. So, $\xi$ values play some inherent role in adjusting the scattering point, hence the polar angle, and finally the beam separation.

In contrast, for the $4.7$ GHz simulation, we choose the scattering point at a lower altitude by setting a very large $\xi$, together with a weaker modulation strength. These constraints produce a weak conal component that remains very close to the core component. Observationally, PSR B1933+16 shows conal-component emergence at $4.7$ GHz and $8.7$ GHz. However, our study indicates that additional components can appear at any rotation phase at sufficiently high radio frequency in the ICS scenario, with the separation solely determined by the dissipation factor $\xi$ and modulation. Therefore, $\xi$ effectively controls the location of the modulation and component spacing via ICS-mediated process. Table \ref{Table1} provides the details of the Gaussian-modulation template used in our simulation.

\begin{table*}
\begin{tabular}{c c c c c c c c c c c} 
 \hline
 Frequency & $\gamma_{sc}$ & ($r_{e}$) & ($\theta_{ip}$) & $\theta_{sc}=1.5\times\theta_{ip}$ & ($\phi^{\prime}$) & $\phi$ & $f_{t1},f_{l1},f_{c}$ & $F_{\phi}$ / $F_{\theta}$ / $F_{\phi^{\prime}}$\\ 
   & & &  & & & & & & &            \\
 \hline
	4.9 GHz & 400 & 1 $r_{Lc}$ & 1$^{\circ}$.35 & 2 & 4$^{\circ}$.3 & 71$^{\circ}$ & 0.92,0.92,1 & $\exp[-(\frac{(\phi-0^{\circ})^{2}}{(0.55)^{2}}]+0.92\exp[-(\frac{(\phi\pm71^{\circ})^{2}}{(0.28)^{2}}]$  \\ 
 \hline
 9 GHz &1200 & * & 9$^{\circ}$.61 & 14$^{\circ}$.41 & 30$^{\circ}$.28 &-- &1,1,1 & $\exp[-(\frac{(\phi^{\prime}-0)^{2}}{(10.5)^{2}}]+1\exp[-(\frac{(\phi^{\prime}\pm30.28)^{2}}{(8.5)^{2}}]$  \\ \hline
\end{tabular}
\caption{Table shows the parameter space used to simulate the multi-frequency behaviour of the Pulsar PSR B1933+16 at 4900 MHz, 9 GHz.$^{*}$ Used generalized formula to evaluate field line constant, and it's treated as a variable as a function of rotation phase. Here $\gamma_{sc}$ denotes the Lorentz factor of the secondary cloud, $r_{e}$ is the field line constant value, $r_{Lc}$ is the light cylinder radius, $\theta_{ip}$ is the polar angle at the intersection point where scattering takes place, $\theta_{sc}$ is the net polar angle after scattering,$\phi^{\prime}$ is the rotational phase, $\phi$ is the magnetic azimuth, $f_{t1},f_{l1},f_{c} $ are the peak magnitude of the modulation at the trailing, leading and core location, $F_{\phi}$,  $F_{\theta}$ ,$F_{\phi^{\prime}}$ denotes the modulation profile set along azimuthal, polar direction and equivalent modulation profile in rotation phase space.
}
\label{Table1}
\end{table*}

\begin{table}

\begin{tabular}{c c c } 

 \hline
 Frequency &$\gamma_{sc}$ & ($r_{e}$) \\
 \hline
	430 MHz GHz & 400 & 26.77 $r_{Lc}$  \\ 
 \hline
 1.4 GHz &400 & 9$r_{Lc}$  \\ \hline
 2.7 GHz &400 & 5.329 $r_{Lc}$  \\ \hline
\end{tabular}
\caption{Table shows the parameter space used to simulate the multi-frequency behaviour of the Pulsar PSR 1933+16 at 430 MHz, 1.4 GHz, 2.7 GHz. The rest of the parameters chosen to simulate the integrated pulse profile are dipole tilt angle $\alpha=44^{\circ}$, line of sight impact angle $\sigma=1^{\circ}$, and distance of the pulsar within $R_{0}=3.87$~Kpc to 4 Kpc. }
\label{Table2}
\end{table}
\begin{table*}
\begin{tabular}{c c c c c c c c c c} 

 \hline
$\nu_{s} $ & $\nu_{i}$ & $\gamma_{p}$ & $r_{sc}$ & $\xi$ & $\gamma=\gamma_{p}\exp[-\xi(\frac{r-R_{Ns})}{r}]$ & $\phi$ & $\phi_{c}$ & $\theta$ & $\theta_{i}$ \\
 \hline
	4.9 GHz & 100 MHz & 10$^{6}$ & 5 $R_{NS}$ & 12.66 & 39.76 & 0 & 180 & 3 & 10   \\ 
 \hline
 4.9 GHz & 100 MHz & 10$^{6}$ & 5 $R_{NS}$ & 12.9952 & 30.55 & 0 & 180 & 3 & 13.029   \\ 
 \hline
  8.7 GHz & 100 MHz & 10$^{6}$ & 5.835 $R_{NS}$ & 11.7905 &57.15 & 0 & 180 & 1$^{\circ}.1$ & 9$^{\circ}.31$   \\ 
 \hline
 8.7 GHz & 100 MHz & 10$^{6}$ & 5.835 $R_{NS}$ & 11.925 &51.12 & 0 & 0 & 1$^{\circ}.1$ & 10$^{\circ}.41$   \\ 
 \hline
\end{tabular}
\caption{Table shows the parameter space to compute the scattered frequency. $\nu_{s}$ represent the scattered frequency, $\nu_{i}$ is the input frequency, $\gamma_{p}$ is the Lorentz factor of the primary plasma, $\xi$ is the dissipation factor or energy loss factor of plasma particles, $\gamma$ is the Lorentz factor of plasma at the scattering point, $\phi$,$\phi_{c}$ are the azimuthal coordinates of two diagonally opposite spark points, located on the circumference of polar cap rim, $\theta$ is the polar angle corresponding to the emission point and $\theta_{i}$ is the angle between radiation direction and local particle direction at the scattering point. The scattering point is a range of areas where secondary charge clouds form in the magnetosphere. Model of Lorentz factor decay can be found in \citet{1997ApJ...491..891Z}.}
\label{Table3}    
\end{table*}
\begin{table}
\begin{tabular}{c c c c c c } 

 \hline
$\nu_{s} $ & $\nu_{i}$ & $\phi_{c}$ & $\xi$ & r &$\phi^{\prime}$ \\
 \hline
	9 GHz&10 MHz & 90$^{\circ}$ &0.7 & 0.01$r_{Lc}$ & 0$^{\circ}$.5  \\  \hline
    9 GHz&10 MHz & 0$^{\circ}$ &0.735 & 0.01$r_{Lc}$ & 0$^{\circ}$.5  \\  \hline
    9 GHz&10 MHz & 93$^{\circ}$ &0.7 & 0.01$r_{Lc}$ & 1$^{\circ}$  \\  \hline
    9 GHz&10 MHz & 0$^{\circ}$ &0.77 & 0.01$r_{Lc}$ & 1$^{\circ}$  \\  \hline
    9 GHz&10 MHz & 95$^{\circ}$ & 0.7 & 0.01$r_{Lc}$ & 1$^{\circ}$.5  \\  \hline
    9 GHz&10 MHz & 0$^{\circ}$ & 0.8 & 0.01$r_{Lc}$ & 1$^{\circ}$.5  \\  \hline
     9 GHz&10 MHz & 99$^{\circ}$ & 0.7 & 0.01$r_{Lc}$ & 2$^{\circ}$.0  \\  \hline
    9 GHz&10 MHz & 0$^{\circ}$ & 0.835 & 0.01$r_{Lc}$ & 2$^{\circ}$.0  \\  \hline
    9 GHz&10 MHz & 102$^{\circ}$ & 0.7 & 0.01$r_{Lc}$ & 2$^{\circ}$.5  \\  \hline
    9 GHz&10 MHz & 0$^{\circ}$ & 0.895 & 0.01$r_{Lc}$ & 2$^{\circ}$.5  \\  \hline
    9 GHz&10 MHz & 104$^{\circ}$ & 0.7 & 0.01$r_{Lc}$ & 3$^{\circ}$.0  \\  \hline
    9 GHz&10 MHz & 0$^{\circ}$ & 0.9 & 0.01$r_{Lc}$ & 3$^{\circ}$.0  \\  \hline
     9 GHz&10 MHz & 105$^{\circ}$.8 & 0.7 & 0.01$r_{Lc}$ & 3$^{\circ}$.5  \\  \hline
    9 GHz&10 MHz & 0$^{\circ}$ & 0.929 & 0.01$r_{Lc}$ & 3$^{\circ}$.5  \\  \hline
    9 GHz&10 MHz & 105$^{\circ}$.8 & 0.7 & 0.01$r_{Lc}$ & 3$^{\circ}$.5  \\  \hline
    9 GHz&10 MHz & 0$^{\circ}$ & 0.929 & 0.01$r_{Lc}$ & 3$^{\circ}$.5  \\  \hline
    9 GHz&10 MHz & 107$^{\circ}$.8 & 0.7 & 0.01$r_{Lc}$ & 4$^{\circ}$.0  \\  \hline
    9 GHz&10 MHz & 0$^{\circ}$ & 0.955 & 0.01$r_{Lc}$ & 4$^{\circ}$.0  \\  \hline
    9 GHz&10 MHz & 111$^{\circ}$.8 & 0.7 & 0.01$r_{Lc}$ & 4$^{\circ}$.5  \\  \hline
    9 GHz&10 MHz & 0$^{\circ}$ & 0.985 & 0.01$r_{Lc}$ & 4$^{\circ}$.5  \\  \hline
    9 GHz&10 MHz & 113$^{\circ}$.5 & 0.7 & 0.01$r_{Lc}$ & 5$^{\circ}$.0  \\  \hline
    9 GHz&10 MHz & 0$^{\circ}$ & 1.01 & 0.01$r_{Lc}$ & 5$^{\circ}$.0  \\  \hline
     9 GHz&10 MHz & 116$^{\circ}$.5 & 0.7 & 0.01$r_{Lc}$ & 5$^{\circ}$.5  \\  \hline
    9 GHz&10 MHz & 0$^{\circ}$ & 1.035 & 0.01$r_{Lc}$ & 5$^{\circ}$.5  \\  \hline
    9 GHz&10 MHz & 118$^{\circ}$.5 & 0.7 & 0.01$r_{Lc}$ & 6$^{\circ}$.0  \\  \hline
    9 GHz&10 MHz & 0$^{\circ}$ & 1.059 & 0.01$r_{Lc}$ & 6$^{\circ}$.0  \\  \hline
\end{tabular}
\caption{Table shows combinations of input parameter space, susceptible to produce scattered frequency at 9 GHz. To generate high frequency radio photons at 9 GHz we have used, a fixed low-frequency photon source at 10 MHz, and different combinations of the input parameters of energy loss factor $\xi$, azimuthal location of sparks on the polar cap rim.}
\label{Table4}    
\end{table}

\section{Discussion}
In the context of radio-pulsar emission theory, the ICS mechanism is regarded as an important physical process, as it provides a natural explanation for the diverse morphologies observed in integrated pulse-profile structures. If one calculates the plasma frequency at a typical magnetospheric height for an electron number density $n_{e} = 10^{12}~\mathrm{cm^{-3}}$, the resulting value is of the order of a few hundred megahertz. Such a high plasma frequency does not permit low-frequency radio waves to propagate freely. However, the ICS models proposed by Qiao and collaborators \citep{1998A&A...333..172Q,2001A&A...377..964Q,2007A&A...465..525Z} assume that low-frequency seed photons are excited by an ad-hoc mechanism, specifically, current oscillations near the vacuum gap close to the stellar surface clearly lagging with the basic foundation of the involvement of the emission mechanism.
Furthermore, the model proposed by \citet{1998A&A...333..172Q} explains different pulse profile morphologies: Type Ia, Type Ib, Type IIa, Type IIb, and Type IIc based on the beam-frequency diagram, i.e., the relation between the emitted frequency and the angle between the radiation direction and the magnetic axis ($\theta_{\mu}$). Their simulations of pulse-profile morphology reproduce Type Ia and Type Ib profiles for core-dominated pulsars, and Type IIa, IIb, and IIc profiles for conal-dominated pulsars. While the conceptual framework of the ICS model and the associated beam-frequency diagram provide valuable physical insight into the variety of pulse-profile shapes, the simulations themselves rely on an imposed Gaussian profile for each component, rather than a simulation of the integrated pulse profile based on first principles. This simplification arises because it is extremely challenging to compute integrated pulse profiles directly from the true source function derived from the radiation of an accelerated plasma ensemble.

In the present model, we simulate the pulse-profile structure by developing a mathematical framework based on curvature radiation produced by the acceleration of individual charged particles. We then apply the standard ICS-based frequency–upscattering formula to low-frequency radio photons. These low-frequency photons deviate from their unperturbed trajectories when they scatter off the secondary plasma clouds. For a dipolar magnetic field, the scattering angle can be written as $\theta_{sc}=(\theta/2)$, where $\theta$ is the polar angle of the original trajectory. Such scattering generates conal emission components in the upshifted frequency domain, which add to the intrinsic core emission and thus contribute to the formation of the integrated pulse profile. A key parameter in our model is the dissipation factor $\xi$, which governs the rate at which the energy of the primary plasma decays into particles with low Lorentz factors (typically $\gamma \sim 10$). If the energy dissipates rapidly, the secondary cloud forms close to the neutron-star surface. Conversely, for a slowly decaying energy profile, the secondary cloud is established farther out in the magnetosphere. Since particles are constrained to follow diverging dipolar field lines, the polar angle increases with altitude for a given field-line constant. Therefore, both the scattering angle and the resulting location of the conal beams in the shifted frequency space depend sensitively on the dissipation factor in the Lorentz-factor decay model, i.e., $\gamma=\gamma_{0}\exp[-\xi (r-R_{Ns})/R_{Ns}]$. In summary, our model presents a self-consistent mathematical treatment that simulates pulse profiles based on single-particle acceleration. At low radio frequencies, a single core component is produced entirely by curvature radiation. At higher radio frequencies, however, the intrinsic core emission becomes convolved with the ICS-generated conal components, thereby forming the full integrated pulse-profile structure.

We template a Gaussian-modulated profile due to ICS scattering, where we set the peak location of the modulation at $\pm\theta_{sc}$ in the polar direction. Then, the Gaussian template for a given strength is superposed with the original source function derived based on the acceleration model, and finally integrated over the emission region across each open field line for a given dipolar tilt angle to estimate the simulated profile. We have shown basically a Type Ia type profile here, which shows a single component at 408 MHz, 1.4 GHz, and 2.7 GHz, but at 4.9 GHz and 9 GHz, it's showing three-component profiles. Component separation and amplitude can be adjusted by setting the $\theta_{sc}$ and peak magnitude of the Gaussian template. However, $\theta_{sc}$ exclusively depends upon the factor $\xi$, which decides the scattering point and hence the final polar angle after ICS interaction.
Currently, in our model, we have not assumed the aberration-retardation effect, which is regarded as the most prominent effect in the pulsar magnetosphere, susceptible to distorting the symmetrical distribution of pulse components across the rotation phase. In our subsequent paper, we plan to include the aberration-retardation effect to understand all types of profile morphology, Type 1a, Type 1b, Type IIa, Type IIb, Type IIc, in the context of ICS.

A wide range of modeling strategies has been developed to understand how propagation effects reshape pulsars' pulse-profile morphology. Analytic ray-tracing approaches solve wave-propagation equations through an assumed plasma distribution; for example, \citet{beskin2023triplepulsarprofilesgenerated} showed that O-mode refraction in an inhomogeneous magnetospheric plasma can naturally reproduce multi-component profiles, including triple peaks. Polarization transfer models instead integrate mode coupling through the magnetosphere, often invoking a polarization-limiting radius to predict full Stokes parameters; \citet{10.1093/mnras/stad2271}, for instance, developed a partial coherence model treating the emission as an incoherent sum of two modes with a small coherent term, successfully reproducing complex polarization behavior. Complementing these physics-based approaches are geometric-centric beam models that assume one or more coaxial cones or extended fan-beam structures; \citet{jaroenjittichai2025frequencyevolutionpulsaremission} used wideband data to show that extended fan beams, spanning a range of altitudes and magnetic azimuths, better capture the frequency evolution of many pulsars than single-height cones. Data-driven classification frameworks such as the ToPP graph algorithm \citep{refId0} further demonstrate that profile diversity reflects continuous variations in core/cone emission and viewing geometry.

In our current model, we assume that O-mode is primarily responsible for the source of low-frequency seed photons for ICS-mediated interaction, which gets scattered off before it gets attenuated completely. In principle, X-mode can also lead to ICS scattering if the ray manages to undergo sufficient refraction and intersect with a secondary cloud zone,  located radially outward. However, there is no conclusive and clear evidence from the observation point of view to understand which particular mode is solely responsible for ICS-mediated interaction for new component generation at a higher radio counterpart in the pulsar magnetosphere. So, it's very difficult to discriminate the type of modes of the low-frequency photon, but for efficient generation of the new component at a higher radio counterpart, ICS interaction must occur with secondary clouds, preferably having a very low Lorentz factor around 10.
To summarize the role of propagation, we stress that neither emission geometry alone nor propagation alone suffices to explain the full richness of observed pulse profiles; a combination of both is essential. Even if the emission mechanism produces a relatively simple intrinsic beam pattern, magnetospheric propagation will substantially modify it. In practice, propagation can broaden or split components, shift their phases, or introduce additional peaks and polarization features that would not arise in a vacuum-emission scenario. Although propagation effects generally weaken at higher radio frequencies (above $\sim 1$ GHz) because the refractive-index contrast between O- and X-modes decreases, they do not vanish entirely. Residual effects may persist in regions of enhanced plasma density or near the magnetic axis. Thus, mapping observed pulse-profile components to their physical origins requires accounting for both the intrinsic emission beam and the subsequent propagation that reshapes the final observed structure. 

\section{Conclusions}
We decipher the following salient features of the model below:
\begin{enumerate}
  \item We have developed a mathematical formulation combining the emission geometry of a single charged particle acceleration mechanism, Gaussian modulation, and inverse Compton scattering of low-frequency photons to predict the simulation behavior of multi-frequency evolution of the integrated pulse profile.

  \item We present the multi-frequency evolution of PSR B1933+16, including the appearance of new components at high radio frequencies.
  %We have clearly shown the simulation of multi-frequency behavior and new component emergence of the pulsar PSR B1933+16 at higher radio frequency.

  \item The beam frequency diagram of PSR B2111+46 shows multiple bounding curves whose crests and troughs delineate the core–cone structure. %We have shown a beam frequency diagram of PSR B2111+46, which shows that the beam frequency diagram is bounded by distinct multiple curves with multiple crests and troughs, clearly demarcating the core-cone structure.

  \item The beam frequency diagram of PSR B1933+16 is not encompassed by multiple curves; rather, it is bounded by two sets of curves, implying that the pulsar has only a single natural component.%Beam frequency diagram of PSR B1933+16 is not encompassed by multiple curves; rather they are bounded by two sets of curves, which implies that the pulsar has only one single natural component.
  \item From the model, it's clear that the net scattering cross-section gets enhanced for a lower value of the Lorentz factor, but for ICS interaction with highly energetic particles, the scattering cross-section falls drastically.

  \item Calculation shows (see \ref{AppendixA}) that higher gap height, higher magnetic field, and higher emission cross section of ICS, reduce the power emissivity ratio ($\eta$), i.e., ratio of curvature radiation and ICS power, but for coherent cases, curvature radiation emissivity factors get enhanced by a significant amount. Other controlling factors of the emissivity ratio $\eta$ are the Lorentz factor, spin period, and emission height; pushing towards higher limiting values of these parameters enhances the $\eta$ factor significantly.
   
\end{enumerate}

\section*{Acknowledgements}
Tridib Roy greatly acknowledges the support of the funding agency: National Science Center, Poland, grant no. 2023/49/B/ST9/01783. Mayuresh Surnis acknowledges IISER Bhopal (MHRD-regulated, Government of India-funded institute) for providing research infrastructure and financial remuneration.

\appendix

\section{Efficiency of curvature radiation and luminosity of ICS mechanism}\label{AppendixA}
The energy loss of a particle through the curvature radiation process is given by the Larmor formula as
\begin{equation}\label{A1}
\dot{E}_{CR}=P_{CR}=\frac{2 q^{2} c \gamma^{4}}{3 \rho^{2}},
\end{equation}
 where $\gamma$ is the Lorentz factor of the particles, q is the electric charge, c is the light velocity, $\rho$ is the radius of curvature of the magnetic field. The curvature radius $\rho$ for a dipole magnetic field generally scales as $ \rho\approx 10^{8} ~P^{1/2}~cm$, where $P$ is the spin period of the Pulsar. The energy loss of the particle through ICS is given by:
\begin{equation}\label{A2}
 p_{ics}=c \sigma_{sc} n_{ph0} \hbar (\omega^{\prime}-\omega_{0}), 
\end{equation}
where,
\begin{equation}\label{A3}
\omega^{\prime}=2\gamma^{2}\omega_{0}(1-\beta\cos\theta_{i}),
\end{equation}
$\hbar = h/2\pi$ with $h$ being the Planck constant,  $\sigma_{sc}$ is the scattering cross-section of ICS, $n_{ph0}$ is the photon number density near the surface, and $\hbar \omega^{\prime}$ is the energy of the outgoing photons near the surface.  For $\theta_{i}=\pi/2$, $\omega^{\prime}\approx 2\gamma^{2}\omega_{0}$. Adjacent to the neutron star surface, the photon number density of the low-frequency wave can be written as:
\begin{equation}\label{A4}
n_{ph0}=\frac{s}{c \hbar \omega_{0}},
\end{equation}
and the pointing flux translates as
\begin{equation}\label{A5}
s=\frac{c}{4\pi}E_{gap}^{2}.
\end{equation}
Here, $E_{gap}$ corresponds to the gap electric field, which is given by \citet{1975ApJ...196...51R} as:
\begin{equation}\label{A6}
E_{gap}=2 \frac{\Omega B H}{c},
    \end{equation}
where $\Omega$ is the rotational frequency of pulsar, B is the magnetic field of pulsar, $H$ is the thickness of the gap, and $r_{p}=R_{NS}\theta_{p}$ is the radius of the Gap, $\theta_{p}$ is the angular radius of polar cap given by $\theta_{p}=\theta_{c}=(2\pi R  / P c)^{1/2}$.
Power radiated in the ICS process near the surface of a neutron star is given by 
\begin{equation}\label{A7}
p_{ics}=\frac{2 B^{2} \gamma^{2} H^{2} \Omega^{2} \sigma_{sc}}{c \pi}
\end{equation}

The ratio $\eta$ of the energy loss because of the curvature radiation to the ICS mechanism, for the generalized case where $n_{ph}=n_{ph0}(R/r)^{3}$, is given by
%The ratio $\eta$ of these two processes near the surface of the neutron star, but for the generalized case we should substitute $n_{ph}=n_{ph0}(R/r)^{3}$, which gives:
\begin{equation}\label{A8}
\eta=\frac{p_{cr}}{p_{ics}}=\frac{3\pi c^{2} q^{2} \gamma^{2} r^{2}}{8 R^{3} B^{3} H^{2} \Omega^{2} R_{c}\sigma_{sc}}\approx 10^{-5} \gamma_{3}^{4} P  r_{8}^{2} B_{12}^{-2} H_{3}^{-2},
\end{equation}
where $\gamma_{3}=\frac{\gamma}{10^{3}}$, $r_8=\frac{r}{10^{8}}~Cm$, $B_{12}=\frac{B}{10^{12}}$ Gauss and $H_{3}=\frac{H}{10^{3}}~Cm$, in details.

The ICS cross section $\sigma_{sc}$ for the low frequency response is given, by \citet{1985A&A...152...93X}:
\begin{eqnarray}\label{A9}
\sigma_{sc}(1) &=&\sigma(1\rightarrow1^{\prime})+\sigma(1\rightarrow2^{\prime}) \nonumber \\
&=&Z_{\beta}\sigma_{th}\left[\sin^{2}\theta_{i}+0.5\left(\left(\frac{\omega}{\omega+\omega_{B}}\right)^{2}+\left(\frac{\omega}{\omega-\omega_{B}}\right)^{2}\right) \right], \nonumber \\ 
\end{eqnarray}
\begin{eqnarray}\label{A10}
\sigma_{sc}(2)&=&\sigma(2\rightarrow1^{\prime})+\sigma(2\rightarrow2^{\prime}) \nonumber \\
&=& 0.5\sigma_{th} Z_{\beta}\left(\left(\frac{\omega}{\omega+\omega_{B}}\right)^{2}+\left(\frac{\omega}{\omega-\omega_{B}}\right)^{2}\right),
\end{eqnarray}
where $Z_{\beta}=(1-\beta\cos\theta_{i})$, 1(1$^{\prime})$ represent the linear polarization of the incoming(outgoing) photon parallel to the plane of magnetic field lines, whereas 2(2$^{\prime})$ represent the linear polarization mode of the same quantities for photons emitted perpendicular to the plane of magnetic field lines. Here $\omega_{B}=(q B)/(m_{e}c)$ is the cyclotron frequency, for $\omega\ll \omega_{B}$, $\sigma_{sc}(1)\gg \sigma_{sc}(2)$. So approximately, we can write
\begin{equation}\label{A11}
\sigma_{sc}(1)=\frac{\sin^{2}(\theta_{i})}{\gamma^{2}Z_{\beta}}\sigma_{th},
\end{equation}
and for $\theta_{i}=\pi/2$ case, we can write 
\begin{equation}\label{A12}
\sigma_{sc}=\sigma_{sc}(1)\approx \gamma^{-2}\sigma_{th}.
\end{equation}
Here $\sigma_{th}$ is the Thomson scattering cross-section. We finally submit the expression of $\sigma_{sc}$ in the $\eta$ expression and get:
\begin{equation}\label{A13}
\eta=1.9\times 10^{-5}\gamma_{3}^{4} P r_{8}^{2} B_{12}^{-2} H_{3}^{-2}   
\end{equation}
So, it is evident that at a few hundred or thousands of kilometres above the surface, ICS is more efficient than Curvature radiation. But coherent curvature radiation can significantly enhance the efficiency factor.
The incoherent Luminosity by the ICS process near the surface of the neutron star is
\begin{equation}\label{eq:Lincoh}
L_{\rm ics,incoh}=\sigma_{sc} c n_{ph}\hbar \omega^{\prime}\tau \frac{d N}{dt},   
\end{equation}
where $d N / dt=\chi \pi r_{p}^{2} n_{0} c$, $\chi$ is a constant, $\tau$ is the characteristic time over which the particle radiates at some desirable frequency. Here $n_{0}$ is the Goldreich Julian charge density given by
\begin{equation}\label{A15}
n_{0}=n_{\rm GJ}\approx \frac{\vec{\Omega}.\vec{B}}{2\pi c q}
\end{equation}
Substituting the $n_{0}$ value in equation (\ref{eq:Lincoh}), the incoherent Luminosity is
\begin{equation}\label{A16}
L_{\rm ics,incoh}=(0.76\times 10^{29}~{\rm erg/s})~\zeta B_{12}^{3}\gamma_{3}^{2}H_{3}^{2}P^{-4}(\sigma/\sigma_{th})
\end{equation}
at $r\gg R$ and choosing $R=100$ km, $\gamma=1000$, the typical luminosity of ICS will be down by 9 magnitudes. The luminosity observed in radio pulsars is given by \citet{1975ApJ...196...51R}:
\begin{equation}\label{A17}
    L=(3.3 \times 10^{25} {\rm erg /s})S_{400}~d^{2},
\end{equation}
where $S_{400}$ and $d$ are the flux at 400 MHz and distance in kilo-parsec. The typical range of $S_{400}~ d^{2}$ is from 10 to 10$^{5}~{\rm mJy-kpc^{2}}$. This clearly implies that an incoherent ICS process alone is insufficient to explain the high brightness temperatures of radio pulsars and points to the need for a strong emission mechanism, such as coherent curvature emission.
Net scattered power in the output radiation due to ICS mediated mechanism is given in equation (7.16a) by \citet{1979rpa..book.....R}:
\begin{equation}\label{A18}
    \frac{d E_{rad}}{dt}=\frac{4}{3}\sigma_{T} c \gamma^{2}\beta^{2} U_{ph}.
\end{equation}
Where $\sigma_{T}=(8\pi/3)r_{0}^{2}$ is the Thomson scattering cross-section, $r_{0}$ is the classical electron radius, $U_{ph}=\int E g(E) dE$ is the photon energy density, and number of photons is $n_{ph}=\int g(E) dE$.
Substituting equation (\ref{A6}) into equation (\ref{A5}), thereafter plugging in the value of $E_{gap}$ into equation (\ref{A4}), we get photon number density:
\begin{equation}\label{A19}
n_{ph0}=\frac{\Omega^{2} B^{2} h^{2}}{\pi E_{gap} c^{2}}
\end{equation}
Therefore, photon energy density roughly translates as:
\begin{equation}\label{A20}
    U_{ph}=E_{gap} n_{ph0}=\frac{\Omega^{2} B^{2} h^{2}}{\pi c^{2}}.
\end{equation}
Further substituting the photon energy density in equation (\ref{A18}) we get the scattered power due to ICS as:
\begin{equation}\label{A21}
\dot{E}_{rad}=\frac{4\sigma_{T} \gamma^{2}\beta^{2} B^{2} \Omega^{2} h^{2}}{3 \pi c}
\end{equation}
So, the ICS cooling rate is given by:
\begin{equation}\label{A22}
    t_{ics}=\frac{\gamma m_{e} c^{2}}{\dot{E}_{ics}}=\frac{3 \pi m_{e} c^{3}}{4\sigma_{T} \Omega^{2} B^{2} h^{2}}.
\end{equation}
For curvature radiation cooling time-scale calculation we need to substitute radius of curvature in equation (\ref{A1}) for the curvature radiation limited spectrum as (see \citet{1975ApJ...196...51R}):
\begin{equation}\label{A23}
\omega_{cr}=\frac{3 \gamma^{3} c}{2 \rho},
\end{equation}
where $\omega_{cr}$ is the angular radiation frequency as predicted by curvature radiation.
So, equation (\ref{A1}) translates as:
\begin{equation}\label{A24}
    \dot{E}_{CR}=\frac{2 q^{2} c \gamma^{4} 16 \pi^{2} \omega_{cR}^{2}}{9 c^{2}\gamma^{6}}=\frac{32 \pi^{2} q^{2}\omega_{cR}^{2}}{9 c \gamma^{2}}
\end{equation}
So, curvature radiation-limited cooling time scale translates as:
\begin{equation}\label{A25}
t_{cr}=\frac{\gamma m_{e} c^{2}}{\dot{E}_{CR}}=\frac{27 m_{e} c^{3} \gamma^{3}}{8 q^{2} \omega_{cr}^{2}}
\end{equation}
Finally, the ratio of cooling time-scale is estimated as:
\begin{equation}\label{A26}
\frac{t_{ics}}{t_{cr}}=\frac{\dot{E}_{cr}}{\dot{E}_{ics}}=\frac{P_{cr}}{P_{ics}}.
    \end{equation}
It is evident from equation (\ref{A8}) that, $P_{ics}>P_{cr}$, implies that $t_{ics}<t_{cr}$. So ICS-mediated photon cools faster than curvature radiation. But when the system emits coherently, $P_{cr}$ becomes comparable to that of $P_{ics}$, in that case cooling time-scale also becomes comparable to each other.

\bibliographystyle{elsarticle-num-names} 
%\bibliography{references_ICS}



%% For example:

%% \bibitem[Aladro et al.(2015)]{Aladro15} Aladro, R., Martín, S., Riquelme, D., et al. 2015, \aas, 579, A101

%%\end{thebibliography}

\end{document}